\documentclass[10pt,a4paper, twocolumn, notitlepage]{article}
\usepackage{rotating}
\usepackage{amsfonts,longtable}
\usepackage{journals}
\usepackage[authoryear,nonamebreak]{natbib}

\def\Ms{$\textrm{M}_{\odot}$}
\def\MspcII{$\textrm{M}_{\odot}/\textrm{pc}^{2}$}

\def\kpc{{kpc}}

\def\H2{H$_{2}$}
\def\HI{H{\sc i}}
\def\HII{H{\sc ii}}
\def\roH2{$\rho_{\textrm{H}_2}$}

\def\MH2{M$_{\textrm{H}_2}$}

\def\M{M\,}

\def\N{N\,}
\def\Malin{Malin\,}

\begin{document}

\title{Can molecular clouds live long?}
 \author{A. Zasov$^{1, 2}$ and A. Kasparova$^{1}$\\
 \\
$^1$ Sternberg Astronomical Institute, Moscow Lomonosov State University\\
$^2$Department of Physics, Moscow Lomonosov State University
 }
 \date{}
 
 \maketitle

\begin{abstract}
It is generally accepted that the lifetime of molecular clouds does not exceed $3\cdot 10^7$~yr due to disruption by stellar feedback.
We put together some arguments giving evidence that a substantial
fraction of  molecular clouds (primarily in the outer regions of a disc) may avoid destruction process for at least $10^8$~yr or even longer.
A molecular cloud can live long if massive stars are rare or absent.
Massive stars capable to destroy a cloud may not form for a long time if a cloud is low massive, or stellar initial mass function is top-light, or if there is a delay of the beginning of active star formation. 
A long duration of the inactive phase of clouds  may be reconciled with the low amount of the observed starless giant molecular clouds 
if to propose that they were preceded by slowly contraction phase of the magnetized \textit{dark gas}, non-detected in CO-lines.
\end{abstract}


\section{Introduction}
Molecular gas in our Galaxy and other spiral galaxies presents a significant fraction of interstellar medium. 
It is the most dense component of gas observed both in a diffuse and a cloudy forms, which is responsible for star formation. 
Hence, the conditions of formation and dissolution of molecular clouds determines the character of evolution of starforming galaxy. 

Molecular clouds (MCs) may be formed either as the result of collision of gaseous flows (the expanding envelopes, supersonic turbulence, large-scale density waves) or by gravitational (magnetogravitational) instabilities of interstellar medium. In the former case the resulting clouds are non-bound as a whole, hence their lifetime is relatively short. 
In the latter case the gravitationally bound clouds {are formed}, including massive giant molecular clouds (GMCs) which  are not so easy to disrupt. The relative role of these two ways of formation is the matter of debate.
Measurements of {cloud} masses and their internal velocity dispersions  indicate that a significant part of MCs 
are virialized (gravitationally bound) or close to the virial state \citep{Heyer2009,Roman-Duval2010,Colombo2014}. On the other hand, according to \citet{Dobbs2011}, many MCs 
are unbound (with the exception of their most dense regions) and short-lived. 
{Note however that the conditions for a clouds to be virialized are badly known, especially if to take into account the magnetic field and surface terms \citep{Ballesteros-Paredes2006}.} Nevertheless the closeness of their masses to the virial threshold evidences the determining role of gravitation in their evolution.  

Usually the term of cloud lifetime implies the period ended by transition of a cloud gas into  diffuse gas phase.
It is widely accepted that even GMCs are short-lived objects ($\le 3\cdot 10^7$~yr), being disrupted by shear motions or by stellar feedback soon after the beginning of star formation due to radiation pressure, stellar winds, SN explosions and the expansion of {\HII} regions \citep[see the discussion of different stellar feedback mechanisms and their efficiencies in][]{Scoville2003, Murray2010, Hopkins2012}. A special case present MCs formed in strong spiral arms of molecular gas~--- rich galaxies. The example of galaxy \M51 shows that giant molecular clouds disintegrate into pieces after passing spiral arm, cross the interarm space and merge again in entering the next arm \citep{Koda2009,Egusa2011}.  In such instances a concept of lifetime of MCs becomes vague, so that formally speaking the life of MCs may be indefinitely long, although the lifetime of a cloud taken as a single whole remains short. 

Some  arguments seem to support the short lifetime of the observed MCs independently on the mechanism of their origin 
\citep[see f.e reviews by][]{Pringle2001,Dobbs2013}. 
First, the molecular clouds are usually observed in the arms of 
spiral galaxies (or in {\HI} filaments in such galaxies as LMC) where they cannot stay
longer than $\sim 10^7$~yr. 
This argument is supported by the fact that star clusters connected with MCs are 
usually younger than $10^6$~yr. Note however, that T\,Tau stars of larger age were also found in the vicinity of some currently
active MCs being drifted from them \citep{Feigelson1996}. There is also a problem of correct comparison of the clouds in the solar neighbourhood with other galaxies. 
As it was pointed out by \citet{Dobbs13}, all relatively nearby molecular clouds which have been age-dated by using HR-diagrams are low-massive. 
Another argument favouring a short timescale is that the overwhelming majority of the observed MCs in our Galaxy and in nearby galaxies reveals the ongoing star-forming activity, so
that the starless MCs are rarely observed \citep{Ballesteros-Paredes2007,Jijina1999,Gratier2012}. It leads to conclusion that star formation begins soon after MC becomes detectable in molecular lines.
However this argument opens 
a question of cloud existence before the active phase of star
formation. 
 
\citet{Murray2011} showed on the statistical ground that the
lifetimes of the observed \textit{massive} ($\sim10^6$~M$_{\odot}$) GMCs of Milky Way characterized by
active star formation is about $27\pm4$~Myr. It agrees with the
lifetime $20-30\cdot10^6$~yr found for MCs in LMC by
\citet{Kawamura2009}. \citet{Dobbs2013} came to similar estimates from the
numerical simulation of the development and dispersion of MCs in spiral arm. However the existing models of formation and evolution of MCs use too many 
simplifications (in particular, a magnetic field is usually ignored).  

In general, there may be two different interpretations of the low frequency of occurrence of
starless MCs. Either their mean lifetime is very short which favours their origin 
in colliding flows or the observed
star-forming clouds do not contain a bulk of molecular gas in the
Galaxy. A growth of gas density during the process of formation of MCs may begin long before they reach their observed density \citep{Dobbs2012}.

In this work we consider the assumption that under certain conditions  molecular clouds may live much longer than $10^7$ yr.  The paper is organized as follows. Section~\ref{longlt} presents an overview of observation facts hinting of the long time scale of clouds. In Section~\ref{survival} we will discuss the possible scenarios which may allow MCs to keep their entity as long as $>10^8$~yr. There  we also show how to reconcile a long survivability of clouds with the low number of the observed starless GMCs.

\begin{figure}[t]
\includegraphics[width=7.5cm]{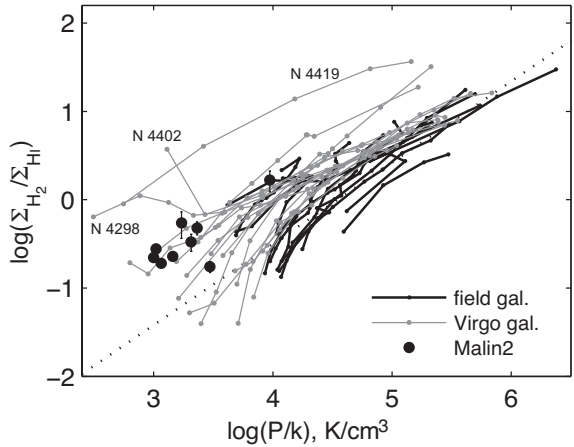}
\caption{Molecular hydrogen fraction versus turbulent gas pressure
in the disc midplane for the field (black lines) and Virgo (gray
lines) galaxies. The black circles denote the areas in the disc of \Malin2.
The dotted line represents the relation proposed for normal galaxies by
\citet{Blitz2006}.}\label{fig1}
\end{figure}

\begin{table}[t]
\footnotesize
\caption{Galaxy sample presented in Fig.~\ref{fig1}. The details can be found in
 \citet{Kasparova2012} and
\citet{Kasparova2014}.} \label{tab}
\begin{tabular}{llllll}
\hline%
\multicolumn{6}{c}{Field galaxies:}
\vspace{1mm}\\
\N628&\N2841&\N3184&\N3198&\N3351&\N3521\\
\N3627&\N4736&\N5055&\N5194&\N6946&\N7331\\
\hline%
\multicolumn{6}{c}{Virgo galaxies:}
\vspace{1mm}\\
\N4254&\N4298&\N4302&\N4303&\N4321&\N4402\\
\N4419&\N4501&\N4535&\N4536&\N4548&\N4567\\
\N4568&\N4569&\N4579&\N4647&\N4654&\N4689\\
\hline
\multicolumn{6}{c}{Giant low surface brightness galaxy \Malin2}\\
\hline%
\end{tabular}
\end{table}

\section{\large Long lifetime of molecular clouds}\label{longlt} 

A possibility of a long lifetime of molecular gas in galaxies has been mentioned earlier by some authors. \citet{Scoville2012} discussed some arguments supporting this idea. The first one is related to a balance condition of transitions between molecular and atomic gas in the interior of galaxies where \H2 dominates in mass compared to \HI. To support this balance the lifetime of a molecular gas may be as large as $10^9$ yr. Similar conclusion that a significant amount of \H2 in molecular gas-rich galaxies remains non-involved in the process of star formation was done by \citet{Kasparova2006} and \citet{Shetty2014}. 
This circumstance raises a question of long-lasted lifetime of \H2. Note however that these arguments consider molecular gas as whole, without dividing it into diffuse and cloud components, so there is no clear evidence that they may be applied to the dense MCs.

Another argument given by \citet{Scoville2012} is that the scaleheights
(perpendicular to the Galactic disc) of the GMCs as a function of their mass
gives evidence that they have achieved approximate energy equipartition. 
This requires the clouds to survive at least several cloud-cloud scattering times ($>10^8$~yr).

Two additional observational evidences of the existence of long-lived MCs (not diffuse gas!) we give below.

\subsection{Absence of the balance \\between {\HI} and \H2}

It is well known that some spiral galaxies in Virgo and other rich clusters are
gas-deficient, especially in their outer parts, as the result of ram pressure of intergalactic medium. 
It is essential, that the \H2-deficieny is less pronounced than \HI-deficiency \citep{Fumagalli2009,Boselli2014}. 
In some CO-emitting galaxies {\HI} is almost completely swept from the external regions of discs, although \H2 remains there  \citep{Corbelli2012,Kasparova2012}. Unlike MCs, diffuse \H2 is stripped by the intracluster gas ram pressure as efficiently as {\HI}, because the higher impact is needed to expel molecular clouds. Numerical high-resolution modellings of gas sweeping process clearly demonstrates this difference \citep[see f.e.][]{Tonnesen2009}. Hence, one can expect that the  observed molecular gas in the outer regions of \HI-stripped galaxies predominantly consists of compact clouds, which are not in a hurry to return back to the atomic state.

For normal non-cluster galaxies there exist the approximate balance between molecular and atomic phases of interstellar gas that manifests itself  as the correlation between the
local ratios of {\H2-to-\HI}  column densities
$\eta\equiv\Sigma_{\rm H_2}/\Sigma_{\rm HI}$ and the gas
turbulent pressure $P$ in the disc midplane \citep[see][and references therein]{Blitz2006,
Kasparova2008}. For a sample of galaxies listed in Table~\ref{tab} we checked   
the balance of gas components
\HI$\longleftrightarrow$\H2 for field galaxies and Virgo members, including the gas-deficient galaxies. To calculate the midplane gas pressure we estimated the
volume densities of disc components assuming its equilibrium and axisymmety, and taking
into account the gravitation of the dark halo and the gas self-gravity, which is especially important for the disc outskirts \citep[see details in][]{Kasparova2012,Kasparova2008}.

Fig.~\ref{fig1} shows the diagram $\eta(P)$ calculated for the field
(black lines) and Virgo spiral galaxies (gray lines). 
In addition to these galaxies, black circles mark the
positions of nine areas in the extended disc of giant low surface
brightness galaxy \Malin2 calculated from the available data  for \H2 in this galaxy by \citet{Kasparova2014}.
As it is clearly seen in  Fig.~\ref{fig1}, the molecular
gas fraction $\eta$ in the periphery of some {\HI}-deficient Virgo galaxies and in the disc of \Malin2 is unusually high for a given gas pressure. 

The ram pressure stripping time scale is about several $10^8$~yr \citep{Vollmer2001,Tonnesen2009}. The question arises why the molecular gas was not dissociated into {\HI} 
shortly thereafter (if the clouds are short-lived), or, if molecules have dissociated, how molecular gas could be formed later under condition of the observed {\HI}-deficiency? In principle, one may admit that the observed molecular gas could persist since the epoch preceding the gas sweeping before they lost their {\HI}, when the galaxies were gas-rich. Indeed, as it was shown by \citet{Kasparova2012}, if to assume that these galaxies possessed {\HI} radial density profiles, typical for field spiral galaxies, the molecular gas ratio $\eta$ for them would also be normal for the existed gas pressure $P$. The lack of active star formation in the outer regions of discs of {\HI}-deficient galaxies may be an essential factor enabling a long lifetime of molecular clouds there.

In the relatively isolated low surface brightness galaxy \Malin2, 
where column gas density is as low as in the peripheries of  high surface brightness galaxies, the situation may be different. 
As it was proposed by \citet{Kasparova2014},  a significant part of gas in this unusual galaxy is probably \textit{dark} (invisible in CO lines and 21~cm), and, hence a gas pressure may be strongly underestimated which explains the obtained high \H2-to-\HI ratios. However there is another possibility, that as in the case of {\HI}-poor Virgo galaxies the observed molecular gas of \Malin2 is not in equilibrium with {\HI}. Being once formed, it lives long in spite of the low gas pressure and low amount of {\HI}.

\subsection{\small Star formation in tidal structures}

Another indication of long life of at least some   MCs is the
existence of star formation sites and young stellar clusters in tidal
tails or debris far from a parent galaxy, and even in intergalactic space in the systems of gas-rich interacting galaxies.
In close vicinity of strongly interacting disc galaxies star formation may be triggered by collisions of stellar/gaseous flows escaping a disc which are responsible for non-linear disturbances \citep[see][and references therein]{Struck2012}. However the emission line regions may be observed at large distances along the tidal structures. Some {\HII} regions are observed even between galaxies, looking as the isolated birthplaces without optically evident connections with
nearby galaxies \citep[see e.g.,][ and references
therein]{Neff2005, Mullan2011, Boquien2009,  Karachentsev2011,
deMello2008, deMello2012}. How such regions, which are evidently related to dense gas, could be formed in the rarefied medium of tidal debris? 

Gravitational  instability  may  explain the formation of massive
gas/stellar condensations ($\sim 10^7-10^9$~\Ms) only in the densest regions of gaseous
tidal tails with \textit{the beads on a string} morphology
\citep{Elmegreen1983, Bournaud2004, Bournaud2008}.
In most cases, the surface density of {\HI} clouds far from parent galaxies is very low \citep[see f.e.][]{deMello2008}, which excludes gravitational instability, especially if we take into account the absence of compressing force of stellar disc gravity. It is natural to assume, that in such cases gaseous clouds or agglomerates, dense enough to be stable against tidal disruption,
existed far before the beginning of star formation. Young stars and {\HII} regions
may form from pre-existing
gas structures in a galactic disc before strong tidal events,
as it was proposed by \citet{Elmegreen1993}.
Seeds of the currently observed star
formation  in tidally disturbed systems may be expelled by tidal
forces together with {\HI} environment from their parent galaxies and,
being gravitationally bound, they avoid the expansion and shear
after leaving a galaxy. In such cases gaseous clumps may be considered as 
long-lived clouds potentially ready for star formation,  the activity of which is delayed up to the timescale of development of tidal structures ($10^8-10^9$~yr).  

Typical line-of-sight column density of {\HI} along a
tidal tail is $10^{20}-10^{21}$~cm$^{-2}$ which corresponds to the
mean volume density $n_H\approx10^{-1}-10^{-2}$~cm$^{-3}$ for the
tail thickness of several \kpc. This density seems too
low for conversion of  significant amount of atomic to molecular gas
within a time of a few hundred Myr. According to \citet{Braine2001}
for a standard distribution of dust grain sizes, for
the transformation of 20 per cent of {\HI} into \H2 on dust grains
it needs time $T_{20}\approx 10^7/n_H$ which exceeds $10^8$~yr. The
background of UV radiation and the reduced dust/gas ratio in the
outskirts of galaxies may only raise this estimate.

Here we pay attention, that  MCs located in tidal
debris, as well as in the outer discs of galaxies, may contain  more
molecular gas than it follows from CO-brightness for the typically assumed conversion factor. It would take place if a gas is extremely cold for excitation of CO emission or if the ratio of CO-to \H2 is very low due to dissolution of CO molecules by UV radiation 
\citep[see][]{Knierman2012,Wolfire2010}. The efficiency of this process depends on dust-to-gas ratio as well as on the density of a cloud, so that the low dense dust-poor MCs may
remain unnoticed. The main source of UV radiation may be a parent galaxy. Indeed, UV photons partially escape from a galaxy especially from its outskirts  due to porosity of interstellar medium. According to \citet{Wolfire2010} a relative fraction of CO-dark molecular gas weakly depends on the background UV intensity, hence UV radiation may be effective both in the periphery and in the environment of a galaxy. The CO-missing gas component is referred to as the
\textit{dark gas}. In our Galaxy the  presence of a dark gas was revealed by
gamma-ray \citep{Grenier2005} and far infrared observations
\citep[see f.e.][]{Pineda2013}. A total mass of this non-luminous gas in the Galaxy is comparable
with the total mass of CO-traced molecular gas,
especially in the outer parts of galactic disc where it prevails.

\section{ \large The survival conditions of clouds} \label{survival} 

The efficiency of the destroying mechanisms for a bound molecular cloud depends on both a cloud structure and the rate of formation of massive stars. Below, we discuss two possibilities for a MC to have a long lifetime, namely (1)~a very low birth probability of massive stars which can disrupt a cloud;
(2)~the delayed star formation.

\subsection{A lack of massive stars}

\subsubsection{Low massive clouds}

 It is difficult to doubt that the intense star formation leads to
 disruption of MCs, when their molecular gas partially
or totally returns to atomic state. Yet the effectiveness of this process is poorly known. The example of spiral galaxy \M51 mentioned above  shows that massive and mostly gravitationally bound  GMCs observed in spiral arms are quite stable against total  destruction by massive stars: they rather disintegrate into pieces than return to atomic diffuse gas. Lower mass MCs may also be stable: they exist  in the interarm space  in spite of continuing star formation there. Later they may collide due to spiral arms streaming motions and merge into giant molecular associations \citep{Egusa2011}.

The persistence of not-too-massive MCs may well be explained by the low probability of formation of a single massive star which can disperse a cloud.
The important factor is the mass of a cloud. 
According to \citet{Williams1997} for MCs with mass $<10^5$~{\Ms} a mean lifetime may
reach $10^8$~yr or more even in the case of continuing
star formation because the massive stars as the principal
agents of cloud destruction are rare. During this time MCs may
easily cross the space between the adjacent spiral arms and
interflow with other molecular aggregates. However, being isolated
(for example, on the disc periphery or in tidal structures), such
{gravitationally bound} cloud may exist till the first massive O-star is formed if a cloud is not destroyed by other forces.

For \citet{Salpeter1955} initial mass function
($\rm dN\propto M^{-\alpha}dM$, $\alpha=2.35$) the formation of a single  massive star ($>10$~\Ms) 
 takes place when the total mass of young stellar population reaches $\rm
 M_{*}\approx200$~\Ms. On the other hand, the ratio of stellar mass formed in GMC to the total cloud mass is $0.03-0.06$ \citep{Evans2009}. Therefore, there is a chance that massive stars do not form if a cloud mass is below $(3 - 7)\cdot 10^3$~\Ms. One may conclude that low-massive clouds which contain a significant fraction of molecular gas in spiral galaxies \citep[see f.e.][]{Koda2009} may survive during the active star formation period, if they are gravitationally bound to resist the shear. It is worth noting that the example of well studied spiral galaxy \M33 shows the growth of fraction of low massive MCs at large radial distances \citep{Gratier2012}. Hence, a chance for MC to avoid disintegration by feedback is higher at the disc periphery.

\subsubsection{Top-light IMF}

Low rate of formation of massive stars and, as a sequence, a long lifetime of MCs may be a result of a top-light initial mass function (IMF)  of stars. Indeed, the universal IMF looks very attractive, because there are no direct evidences that in normal (not star bursting) galaxies IMF is different even for peripheral regions \citep{Goddard2010,Koda2012}. Nevertheless a comparison of IMF of individual young stellar clusters demonstrate that significant IMF variations are real \citep{Dib2014}. Some possible cases of abnormal IMF are discussed by \citet{Saburova2011}. In addition, \citet{McWilliam2013} found that stellar abundance of low density nearby  dSph galaxy (Sgr) reveals the IMF deficiency in the highest mass stars. Another example give LSB galaxies where a steep IMF may explain the combination of metallicity and color indices \citep{Lee2004}. The different behaviour of UV and H$_\alpha$ radial profiles of spiral galaxies also suggests the variation of IMF along radius  \citep{Boissier2007,Krumholz2008,Meurer2009}, although it also may be explained for the standart IMF \citep{Goddard2010}.

As it was found by Hershel observatory for star forming regions  of the Gould Belt, mass function of prestellar cores closely follows a stellar IMF \citep{Andre2010}, 
hence one can expect that the upper end of IMF depends on the properties of dense cores inside MCs, such as their density profile. In general, the upper end of IMF is sensitive to 
the interstellar gas density and/or  star formation rate (SFR), being top-heavy for the higher pre-cluster cloud core density \citep{Marks2012} or high integrated SFR  \citep{Weidner2013,Kroupa2013}. Numerical simulations demonstrate that massive stars are much less likely to form in flat density distributions of turbulent cloud cores \citep{Girichidis2011}. According to \citet{Krumholz2008}, a column density of protostellar clouds should be 
high enough ($\sim1$~g\ cm$^{-2}$) to avoid fragmentation and form massive stars. \citet{Weidner2010} presented observational evidence for a physical relation between a maximal 
stellar mass and total mass of stellar cluster. All these circumstances may be responsible for the environmental variation of the IMF upper end, and, as a result, for the ability of certain part of MCs to survive during star formation.
  
Note that it is hard to distinguish between the variations of the upper limit of stellar masses and the steepness of IMF. 
If to assume $\alpha>3$ and the  total ratio of mass of born stars to the cloud mass $\sim0.05$, then there is a high probability that neither massive star ($>10$~\Ms) forms in the cloud with mass $<10^5$~{\Ms} during the period of star formation. 
Hence, the cloud lifetime will be an order of magnitude longer than for the Salpeter' IMF ($\sim 10^8$~yr). How often it takes place in reality is the open question.

\subsection{Delay of star formation}

The other way to provide  a longevity of MCs is to assume that 	there is a mechanism which inhibits a fast cloud contraction and delays the beginning of 
formation of dense cores and stars. The problem is how to reconcile the delayed star forming activity of  MCs  with a 
 small fraction of observed starless
massive clouds. For example, one may propose that  before the formation of the observed  CO-emitting cloud its molecular gas 
is predominantly in a stage which is non-traceable in CO-lines.  As it was pointed out by \citet{Pringle2001}, if GMCs are formed from molecular gas, the observed clouds may just represent the regions in the ISM
where the molecular gas becomes detectable, evidently because  it is heated by the new-born stars. Molecular gas may  miss a direct detection not only due to extremely low temperature, but also because of dissociation of CO-molecules by UV background radiation, whereas \H2 survives photodissociation. Far infrared observations revealed that the total amount of dark molecular gas in our Galaxy is comparable with the CO-visible gas, especially in the periphery of the Galaxy, where the gas density and dust abundance is lower \citep{Plank2011, Pineda2013}. The analysis of Planck data and {\HI} \citep{Plank2011,Fukui2014} gave evidence that the non-emitting dark gas may also include a large amount of cold and optically thick {\HI}, non-detected in \HI-line. 

Due to UV radiation, CO molecules do not reveal themselves in the outer regions of MCs until the optical
absorption becomes $A_V>3$ while \H2 may exist at $A_V\sim 1$
\citep{Hollenbach1997,Wolfire2010}. A threshold $A_V\sim 3-4$ is consistent with other
findings, both from gamma-ray observations \citep{Ackermann2012} and
from computer simulations \citep{Glover2010}. Hence, one may assume
that the optical thickness of the dark gas in the outer layer of
 a cloud $\Delta A_V\sim 2.5-3$. Using a standard
relationship between the absorption and the column density of
hydrogen 
$$A_V \approx N_H\cdot Z/2\cdot10^{21}$$
where $Z$ is the gas abundance in solar units, one can obtain that
the expected column density $N_H$ of CO-deficient layer for $Z<Z_{\odot}$ is about
$2\cdot10^{22}$~cm$^{-2} \approx 10^2 $~{\MspcII} which is close or even higher than 
typical values of the average surface density CO-observed
MCs \citep[e.g.,][]{Meyer2013}. It illustrates that a
significant mass fraction of molecular clouds may consist mostly  of 
a dark gas which shields its inner CO-emitting gas. It agrees with the data obtained by the Hershel observatory, detected
the [C{\sc ii}]-line emission in the regions of warm and diffuse CO-dark \H2 \citep{Pineda2013}, and Planck
Observatory \citep{Plank2011} confirming that the masses of
CO-emitting and CO-silent molecular gas are comparable. 

Relatively high ionization of the rarefied dark gas containing [C{\sc ii}] ions makes the ambipolar
diffusion very ineffective. Hence, for clouds containing the dark gas ready to collapse a presence of magnetic field strongly retards the contraction. So the time interval between 
the rarefied state of a cloud at the beginning of its formation  and the appearance of CO-loud inner structure
may be very long in comparison with the free-fall time. 

Note that magnetic field evidently plays a crucial role even at later stages of evolution of MCs. The densities of magnetic, turbulent and gravitational energies in the
observed MCs are usually comparable \citep[see e.g.][]{Myers1988},
so that in many cases the
magnetic energy clearly prevails over other types of internal energy types \citep{Giannetti2013, Heyer2009}. The discovery of non-random geometry of
magnetic field inside of clouds found in \M33 and in our Galaxy 
confirm  that the magnetic field energy may dominate over turbulent
energy and the energy of rotation \citep{Li2011, Li2014}.
All these facts give evidence that a magnetized MCs might remain subcritical or at least critical 
since their formation as a bounded single entity. 

Following \citet{McKee1993},
we adopt the analytical expression for the ambipolar diffusion
time
$$t_{AD} \sim 1.6\cdot 10^{14} \cdot X_i\ \mbox{(yr)}$$
where $X_i$ is the gas ionization, determined by balancing the rate
of the ion formation with the rate of their recombination. In turn,
$$X_i\propto(F_{rad}/n_H)^{0.5}$$
where $F_{rad}$ is the ionizing flux and $n_H$ is the atom number
density. For the typical density $n_{H_2} \approx 10^3$~cm$^{-3}$,
the ionization $X_i$ is between $10^{-5}$ and $10^{-7}$, the lowest value
being for those parts of MCs ($A_V>4$) which are shielded from UV
radiation \citep{McKee1993}. According to \citet{Wolfire2010}
for the dark molecular envelope of MC $X_i\sim 10^{-4}$ which 
gives  $t_{AD} \sim 10^9$~yr. It confirms that the gravitational contraction of a subcritical cloud, whose density is lower than the typical value
$n_H$,  may be slowed down by magnetic tension for a long time interval.

\section{Summary}

We put together several arguments supporting the idea that at least a
fraction of {gravitationally bound} molecular clouds may live much longer that it is usually
accepted (up to $\sim 10^8-10^9$~yr).
In particular, it may explain a high \H2 over {\HI} ratio in the peripheries of \HI-deficient Virgo galaxies experienced a ram pressure, as well as the observed star formation in the tidal
structures and in the intergalactic space far from parent galaxies. Most probably, in these cases molecular gas is not in equilibrium with {\HI}.

We considered two possibilities for MCs to have a long lifetime: 
(1)~a low probability of formation of massive stars, capable to destroy a cloud, if a cloud mass is low ($<10^4$~\Ms) or stellar IMF is  top-light, and (2)~ a delayed 
star formation in MCs. The latter scenario may be reconciled with the small amount of the observed  starless GMCs if to propose that  molecular clouds  slowly formed from the magnetized CO-silent (dark)  gas before they become detected in CO-lines. This picture agrees with with the observations revealing a large amount of dark gas in the Galaxy, especially at large radial distance.

\textbf{Acknowledgements.}

We are grateful to Igor Chilingarian for fruitful discussion.
This work was supported by the Russian Foundation
for Basic Research (proj. 12-02-00685).

\bibliographystyle{aa}
\bibliography{lifetime}

\end{document}